\documentclass[12pt]{article}

\usepackage[height=8.85in,width=6.45in]{geometry}

\usepackage[utf8]{inputenc}
\usepackage[T1]{fontenc}
\usepackage{amsmath}
\usepackage{amssymb}
\usepackage{mathtools}
\numberwithin{equation}{section}
\allowdisplaybreaks

\usepackage{slashed}
\usepackage{braket}
\usepackage[usenames,dvipsnames,svgnames,table]{xcolor}
\usepackage[colorlinks,citecolor=DarkGreen,linkcolor=FireBrick,urlcolor=FireBrick,linktocpage]{hyperref}
\usepackage{cite}
\usepackage{graphicx}
\usepackage{times}
\usepackage[scaled]{couriers}

\usepackage{bm}
\usepackage{subfig}

\usepackage{tikz,pgf}
\usepackage{tikz-cd}
\usetikzlibrary{shapes}
\usetikzlibrary{calc}
\usetikzlibrary{decorations.pathmorphing}
\usetikzlibrary{decorations.pathreplacing,shapes.misc}
\usetikzlibrary{positioning}
\usetikzlibrary{decorations.markings}
\tikzset{->-/.style={decoration={
  markings,
  mark=at position .5 with {\arrow{>}}},postaction={decorate}}}
\tikzset{-<-/.style={decoration={
  markings,
  mark=at position .5 with {\arrow{<}}},postaction={decorate}}}

\usepackage{xcolor}
  \definecolor{rblue}{RGB}{81, 49, 193}
  \definecolor{rorange}{RGB}{255, 147, 40}
  \definecolor{rgreen}{RGB}{176, 233, 0}

\renewcommand{\hat}{\widehat}

\usepackage{mdframed}

\begin{document}

\begin{center}

{\large \bfseries Anomaly constraint on chiral central charge of (2+1)d topological order}

\bigskip
\bigskip
\bigskip

Ryohei Kobayashi
\bigskip
\bigskip
\bigskip

\begin{tabular}{ll}
 Institute for Solid State Physics, \\
University of Tokyo, Kashiwa, Chiba 277-8583, Japan\\

\end{tabular}

\vskip 1cm

\end{center}

\noindent 
In this short paper, we argue that the chiral central charge $c_-$ of a $(2+1)$d topological ordered state is sometimes strongly constrained by 't Hooft anomaly of anti-unitary global symmetry. For example, if a $(2+1)$d fermionic TQFT has a time reversal anomaly with $T^2=(-1)^F$ labeled as $\nu\in\mathbb{Z}_{16}$, the TQFT must have $c_-=1/4$ mod $1/2$ for odd $\nu$, while $c_-=0$ mod $1/2$ for even $\nu$. This generalizes the fact that the bosonic TQFT with $T$ anomaly in a particular class must carry $c_-=4$ mod $8$ to fermionic cases. We also study such a constraint for fermionic TQFT with $U(1)\times CT$ symmetry, which is regarded as a gapped surface of the  topological superconductor in class AIII.

\setcounter{tocdepth}{2}
%\tableofcontents

\section{Introduction}
The 't Hooft anomaly in quantum field theory is a mild violation of the conservation law due to quantum effects. It is well known that the 't Hooft anomaly constrains the low energy behavior of the system, since we need nontrivial degrees of freedom in IR to match the given anomaly. 
For example, the seminal Lieb-Schultz-Mattis theorem~\cite{Lieb1961,Hastings,Oshikawa} and its generalizations~\cite{Parameswaran,HarukiCrystal,JianLSM, KobayashiLSM, MengLSM} provide a strong spectral constraints on quantum systems on lattice, which are understood as the consequence of 't Hooft anomaly involving lattice spatial symmetries that behave internally in infrared~\cite{Furuya,ChoManifestation,Cheng_2016LSM, MetlitskiThorngren,yuan,TanizakiCPT,TanizakiSU3}.

The 't Hooft anomaly is typically matched by a symmetry broken or gapless phase (e.g., perturbative anomaly), but in some cases the anomaly is known to be matched by a symmetry preserving gapped phase, realized by Topological Quantum Field Theory (TQFT) enriched by the global symmetry~\cite{Burnell2013Exactly,Fidkowski2014,Wang2013Gapped,Bonderson2013Time,Chen2013Enforced,Metlitski2013Symmetry,Chen2015anomalous,Kapustin2014anomalous,Metlitski2014Interaction,ThorngrenKeyserlingk2015,Witten:2016cio,Wang:2017loc,KOT2019, Kobayashi2019pin,Bulmash_2020}. 
This implies that an anomaly in some particular class can be carried by topological degrees of freedom, not by gapless particles and in particular the system with an anomaly can have an energy gap.
Recently, it was also discovered that some global anomalies cannot be matched by a symmetry preserving TQFT and lead to even stronger spectral constraints~\cite{CordovaOhmori, CordovaOhmorichiral, ThorngrenSSB}.

In this paper, we work on symmetry preserving TQFT with 't Hooft anomaly in $(2+1)$ dimensions, and explore the constraints on the gapped phase enforced by the anomaly. We find that the chiral central charge $c_-$ of the TQFT is strongly constrained by the 't Hooft anomaly of anti-unitary global symmetry.
This can be understood as a constraint on thermal Hall conductance observed on the surface state of a topological superconductor based on time reversal symmetry. 
The result of this paper also implies that the $(2+1)$d topological ordered state with an anomalous $T$ symmetry of a specific index must have a quantized energy current on the $(1+1)$d boundary, which is proportional to the chiral central charge~\cite{Kapustin_2020}.

Here, let us illustrate what we mean by the chiral central charge of a $(2+1)$d gapped phase.
If there is the boundary for the $(2+1)$d gapped phase realized by a $(1+1)$d CFT, we can define the chiral central charge $c_-$ via the chiral central charge of $(1+1)$d CFT on the boundary.
As a canonical example, the gravitational Chern-Simons theory $\mathrm{CS}_{\mathrm{grav}}$ has $c_-=1/2$ on the boundary.
We can also observe the chiral nature of $(2+1)$d gapped theory as a sort of quantum anomaly of the $(2+1)$d gapped phase, even without making up a $(1+1)$d boundary.
Namely, the gravitational Chern-Simons term has the framing anomaly characterized by the bulk topological action $\mathrm{Tr}(R\wedge R)/(192\pi)$, since the gravitational Chern-Simons theory is defined on a spin manifold as
\begin{align}
    \int_{M=\partial W}\mathrm{CS}_{\mathrm{grav}}=\pi\int_W\hat{A}(R)=\frac{1}{192\pi}\int_W\mathrm{Tr}(R\wedge R).
\end{align}
Once we know the anomaly $\mathrm{Tr}(R\wedge R)/(192\pi)$ can be expressed as the gravitational Chern-Simons theory, and once we know $\mathrm{CS}_{\mathrm{grav}}$ has $c_-=1/2$, then we can combine them together to find that $\mathrm{Tr}(R\wedge R)/(192\pi)$ implies $c_-=1/2$. 
Thus we can say that the $(2+1)$d TQFT has the chiral central $c_-$, if the theory has the framing anomaly given by
\begin{align}
    \frac{c_-}{96\pi}\mathrm{Tr}(R\wedge R).
\end{align}

Now we summarize the result of this paper. We start with  time reversal symmetry with $T^2=(-1)^F$ of fermionic TQFT (known as class DIII~\cite{Schnyder_2008, Kitaev_2009, Freed:2016rqq}), whose anomaly is classified by $\mathbb{Z}_{16}$~\cite{Fidkowski2014}. We show that, if the TQFT has a $T$ anomaly labeled by an odd index $\nu\in\mathbb{Z}_{16}$, the TQFT must carry $c_-=1/4$ mod $1/2$, while for even $\nu\in\mathbb{Z}_{16}$, the TQFT must instead carry $c_-=0$ mod $1/2$.

We also consider $T$ anomaly in bosonic TQFT, and show that we must have $c_-=4$ mod $8$ for some particular class of the anomaly, while $c_-=0$ mod $8$ for the other class. This result in the bosonic case is essentially known in~\cite{Barkeshli2016}, but we provide an alternative understanding for this phenomena, which is also applicable for fermionic cases. 
We also study a more involved fermionic TQFT with $U(1)\times CT$ symmetry (known as class AIII), and obtain a constraint $c_-=1/2$ mod $1$ for a specific class of the anomaly regarded as a surface state of a topological superconductor~\cite{WangSenthil}.

\section{SPT phases on time reversal symmetry defects}
Let us consider a $(2+1)$d TQFT with the time reversal symmetry $T$ that suffers from an 't Hooft anomaly. 
In our discussion, we couple the anomalous $(2+1)$d TQFT with a $(3+1)$d SPT phase based on the $T$ symmetry,~\footnote{
In this paper, we don't make a distinction between the SPT phases and the invertible field theories;
what are referred to as SPT phases in the main text should more properly be called as invertible phases.
An invertible phase is a quantum field theory with a unique ground state on an arbitrary closed spatial manifold.
An SPT phase is usually defined as an equivalence class of short-range-entangled gapped Hamiltonian systems with a specified symmetry.
An SPT phase in this sense determines an equivalence class of invertible phases, by isolating its ground state, 
but it is a difficult and unsolved problem whether an arbitrary invertible phase associated to a global symmetry can be realized as an SPT phase in this sense.
Invertible phases also include e.g.~the low-energy limit of the Kitaev wire, which is not counted as an SPT phase in the standard usage in the literature on condensed matter physics,
but is often called as an SPT phase in the high energy physics literature.
} and regard the anomalous $(2+1)$d TQFT as a boundary state of the $(3+1)$d SPT phase.
In a general quantum theory with a global symmetry, there exists a codimension one topological operator that implements the symmetry action. 
We call this topological operator a symmetry defect.
In the $(3+1)$d $T$ SPT phase, let us consider a symmetry defect for the $T$ symmetry, which implements the orientation reversal of the spacetime.

In general, for a $d$-dimensional SPT phase with the $T$ symmetry, the $T$ symmetry defect itself becomes a SPT phase with the unitary $\mathbb{Z}_2$ symmetry.
This phenomena can be understood in the phase where the $T$ symmetry is spontaneously broken. 
Then, the $T$ defect is realized as a $T$ domain wall separating two distinct vacua of the symmetry broken phase.

Concretely, let us consider an infinite system of the $(3+1)$d $T$ SPT phase, and make up a $T$ domain wall of the SPT phase by breaking the $T$ symmetry, dividing the system into the left and right domain. 
We are interested in a theory supported on the $T$ domain wall in this setup. To study the localized degrees of freedom on the domain wall, it is important to ask what the global symmetry of the domain wall is.

\begin{figure}[htb]
\centering
\includegraphics{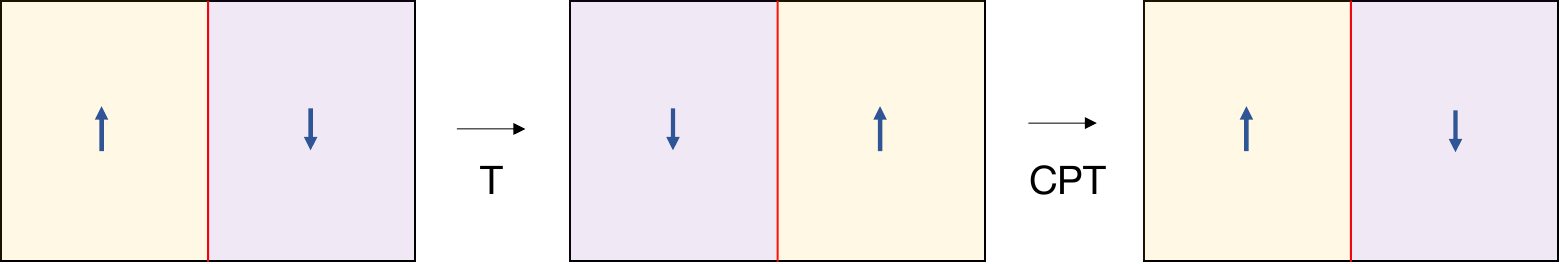}
\caption{The illustration for the $T$ domain wall.
 The $T$ domain wall separates the two distinct vacua in the $T$ broken phase. The $T$ symmetry acts on the figure by changing two vacua (i.e., yellow $\leftrightarrow$ purple). Since the $CPT$ commutes with $T$ (up to fermion parity), the $CPT$ leaves the vacua of the $T$-broken phase invariant, and acts as the parity that reflects the figure across the domain wall.
$T$ alone cannot be a symmetry on the domain wall since it flips the domain, but $T\cdot(CP_{\perp}T)$ works as the symmetry on the wall, since $CP_{\perp}T$ reflects back the configuration of domains.}
\label{fig:smith}
\end{figure}

Throughout the paper, we assume that the theory is Lorentz invariant. In that case, we can find a global symmetry induced on the domain wall, with help of the $CPT$ symmetry~\cite{HKT2019CPT, COSY2019decorated}. 
Concretely, if the $T$ domain wall is located in a reflection symmetric fashion, the combined transformation of $T$ and $CP_{\perp}T$ fixes the domains, and thus acts solely on the domain wall. Here, $P_{\perp}$ denotes a spatial reflection fixing the configuration of the domain wall, see Fig.~\ref{fig:smith}.
Since the $CPT$ is anti-unitary, the combined transformation $T\cdot(CP_{\perp}T)$ turns out to behave as a unitary $\mathbb{Z}_2$ symmetry on the domain wall. The theory on the $T$ domain wall is based on this induced $\mathbb{Z}_2$ symmetry.

In fact, there is a linear relation between the classification of the $(2+1)$d SPT phase on the $T$ symmetry defect and that of $(3+1)$d $T$ SPT phases~\cite{HKT2019CPT, COSY2019decorated}.
This relationship allows us to determine the classification of the $(3+1)$d SPT phase, from a given theory on the symmetry defect. 
This linear map between SPT classifications is nicely formulated in terms of the classification scheme of SPT phases given by cobordism group~\cite{Kapustin:2014dxa}. 
Here, we briefly review the formulation of the map.

First, SPT phases in $(d+1)$ spacetime dimension are classified by the cobordism group $\Omega^{d+1}_{\mathrm{str}}$, where $\mathrm{str}$ stands for spacetime structure that corresponds to the global symmetry, i.e., the choice of internal symmetry and the spacetime symmetry such as fermion parity and/or time reversal~\cite{Kapustin:2014tfa,Kapustin:2014dxa,Freed:2016rqq,Yonekura:2018ufj}. If the structure group is the direct product of the internal symmetry $G$ and the spacetime symmetry, we sometimes write the cobordism group in the form of $\Omega^{d+1}_{\mathrm{spacetime}}(BG)$, where $\mathrm{spacetime}$ denotes the spacetime symmetry.

Then, for a given $(d+1)$d SPT phase with a structure group $\mathrm{str}$ and a codimension one symmetry defect of the $\mathbb{Z}_2$ global symmetry,
we can define the linear map based on the induced structure on the symmetry defect,
\begin{align}
    \Omega^{d}_{\mathrm{str'}}\to\Omega^{d+1}_{\mathrm{str}},
\end{align}
where $\mathrm{str'}$ denotes the structure for the induced symmetry on the symmetry defect. This map of cobordism groups is called the Smith map. 

For example, if we have unitary $\mathbb{Z}_2$ symmetry for the fermionic phase in a $(d+1)$d spacetime $X$, $X$ is equipped with $\mathrm{spin}$ structure on $TX$, and the $\mathbb{Z}_2$ gauge field.
The SPT classification in $(d+1)$d is $\Omega^{d+1}_{\mathrm{str}}=\Omega_{\mathrm{spin}}^{d+1}(B\mathbb{Z}_2)$. 
If we consider the $\mathbb{Z}_2$ symmetry defect $Y$ in $X$, the induced structure on $Y$ from that of $X$ is $\mathrm{spin}$ structure on $TY\oplus NY$, since $TX$ is decomposed into a tangent and normal bundle on $Y$. This structure is shown to be equivalent to $\mathrm{pin}^-$ structure on $Y$. Thus, we have the Smith map
\begin{align}
    \Omega^{d}_{\mathrm{pin}^-}\to\Omega^{d+1}_{\mathrm{spin}}(B\mathbb{Z}_2),
\end{align}
which reflects that anti-unitary symmetry $T^2=1$ is induced on the symmetry defect from unitary $\mathbb{Z}_2$, via the $CPT$ theorem.
The detailed description about the property of the Smith map is found in~\cite{HKT2019CPT}.
In the following discussions, we determine this linear Smith map by considering several cases that span the SPT classification we are interested in.

\subsection{$(3+1)$d bosonic $T$ SPT phase}
\label{subsec:boson}
In the bosonic case, the Smith map determines the classification of $(3+1)$d $T$ SPT phase from that of $(2+1)$d $\mathbb{Z}_2$ SPT phase on the $T$ symmetry defect, expressed as
\begin{align}
    \Omega_{\mathrm{SO}}^3(B\mathbb{Z}_2)\to\Omega_{\mathrm{O}}^4,
    \label{smith:boson}
\end{align}
where $\mathrm{SO}$ and $\mathrm{O}$ denote the oriented and unoriented structure, respectively.
The SPT classification is $\Omega_{\mathrm{SO}}^3(B\mathbb{Z}_2)=\mathbb{Z}_2\times\mathbb{Z}$, and $\Omega_{\mathrm{O}}^4=\mathbb{Z}_2\times\mathbb{Z}_2$. We label the elements of $\Omega_{\mathrm{SO}}^3(B\mathbb{Z}_2)$ as $(n_{DW}, n_{E})\in\mathbb{Z}_2\times\mathbb{Z}$. 
The generators are described as follows:
\begin{itemize}
    \item $(1,0)$ corresponds to the $\mathbb{Z}_2$ SPT phase given by the classical action
    \begin{align}
    \exp\left(\pi i \int a^3\right)
    \label{a3}
\end{align}
with a $\mathbb{Z}_2$ gauge field $a$, which characterizes a nontrivial element of $H^3(B\mathbb{Z}_2, U(1))=\mathbb{Z}_2$.
\item $(0,1)$ corresponds to the $E_8$ state~\cite{Kitaevanyons} with chiral central charge $c_-=8$.
\end{itemize}
Meanwhile, we label the $(3+1)$d $T$ SPT classification by  $(m_{1}, m_{2})\in\mathbb{Z}_2\times\mathbb{Z}_2$, whose generators are described as follows:
\begin{itemize}
    \item 
    $(1,0)$ corresponds to the classical action
    \begin{align}
    \exp\left(\pi i\int w_1^4\right),
    \label{w14}
\end{align}
where $[w_1]\in H^1(M,\mathbb{Z}_2)$ is the first Stiefel-Whitney class of the spacetime $M$.

\item $(0,1)$ corresponds to the classical action
\begin{align}
    \exp\left(\pi i\int w_2^2\right),
    \label{w22}
\end{align}
with $[w_2]\in H^2(M,\mathbb{Z}_2)$ the second Stiefel-Whitney class.
\end{itemize}
The Smith map $\mathbb{Z}_2\times\mathbb{Z}\to \mathbb{Z}_2\times\mathbb{Z}_2$ for~\eqref{smith:boson} is given in the form of 
\begin{align}
(n_{DW},n_E)\to(\alpha_1 n_{DW}+\alpha_2 n_E, \beta_1 n_{DW}+\beta_2 n_E),
\end{align}
for coefficients $\alpha_1, \alpha_2, \beta_1, \beta_2\in \mathbb{Z}_2$. We determine these coefficients by finding what $(2+1)$d phases map to the action~\eqref{w14},~\eqref{w22} respectively.
We will see that $\alpha_1=1,\alpha_2=0,\beta_1=0,\beta_2=1$ in the following discussions.  

We find the theory on the $T$ symmetry defect for~\eqref{w14}, by twisted compactification of~\eqref{w14} with respect to the $T$ symmetry.
It turns out that the restriction of the $T$ defect on the $T$ symmetry defect is regarded as the $\mathbb{Z}_2$ gauge field $a$, and the compactified action is given by~\eqref{a3}. This determines $\alpha_1=1, \alpha_2=0$.

To find the theory on the $T$ symmetry defect for~\eqref{w22}, it is convenient to consider the $(2+1)$d gapped boundary of the SPT phase that preserves the $T$ symmetry. The gapped boundary is realized by the $\mathbb{Z}_2$ gauge theory given by the action
\begin{align}
    \exp\left(\pi i\int a\cup\delta b+a\cup w_2+b\cup w_2\right),
\end{align}
with dynamical $\mathbb{Z}_2$ gauge fields $a,b$. This action realizes a TQFT known as the 3-fermion state~\cite{HsinLorentz, Barkeshli2016, Thorngren_framed}; a $(2+1)$d $\mathbb{Z}_2$ gauge theory whose electric and magnetic particle are both fermions. 
In general, a $(2+1)$d bosonic topological ordered state is described by the fusion and braiding properties of anyons, 
which are characterized by an algebraic theory of anyons known as a unitary modular tensor category (UTMC).
For a given UTMC that describes a $(2+1)$d TQFT, there is a way to compute the chiral central charge $c_-$ modulo 8 known as the Gauss-Milgram formula, given by $e^{2\pi i c_-/8}=\sum_{a}d_a^2\theta_a/\mathcal{D}$~\cite{Kitaevanyons}.
~\footnote{
We can see the correspondence of the Gauss-Milgram formula with the framing anomaly based on the following argument.
Starting from the UTMC $\mathcal{C}$, we can construct the $(3+1)$d Walker-Wang model~\cite{walker2012}, a $(3+1)$d SPT phase whose boundary is given by a $(2+1)$d TQFT described by the UTMC $\mathcal{C}$. Then, it has been shown in~\cite{Barkeshli2016} that the partition function of the Walker-Wang TQFT on the complex projective space $\mathbb{CP}^2$ produces the Gauss-Milgram formula
\begin{align}
    Z(\mathbb{CP}^2)=\frac{1}{\mathcal{D}}\sum_{a}d_a^2\theta_a.
\end{align}
Meanwhile, we can see that
\begin{align}
    \exp\left(\int_{\mathbb{CP}^2}\frac{ic_-}{96\pi}\mathrm{Tr}(R\wedge R)\right)=e^{2\pi ic_-/8},
\end{align}
by recalling that $\mathbb{CP}^2$ has the signature 1. Hence, supposing that the Walker-Wang model is effectively described by the $R\wedge R$ action, the Gauss-Milgram formula exactly computes the framing anomaly $c_-$.
}

Here, the sum is over anyons of the UTMC, $d_a$ is the quantum dimension, $\theta$ is the topological spin, and $\mathcal{D}$ is the total dimension given by $\mathcal{D}^2=\sum_a d_a^2$. According to this formula, we can immediately see that the 3-fermion state has the chiral central charge $c_-=4$ mod $8$.

Then, let us break the $T$ symmetry simultaneously in the $(3+1)$d bulk and the $(2+1)$d boundary, such that the $T$ domain wall in the bulk terminates on a domain wall at the boundary. 
The $T$ domain wall separates the left and right domain, see Fig.~\ref{fig:wall}. Let us assume that one specific realization of our system has $c_- = 4+8m$ for $m\in\mathbb{Z}$, on the boundary of the left domain. 

Since the right domain can be prepared as a partner of the left domain conjugate under the $T$ symmetry, the boundary of right domain carries $c_-=-(4+8m)$, since the orientation at the right domain gets reversed by the $T$ action. This implies the $(3+1)$d SPT action given by
\begin{align}
    \frac{c_-}{96\pi}\mathrm{Tr}(R\wedge R)
    \label{eq:bulkRR}
\end{align}
has a kink of $c_-$ from $c_- = 4+8m$ to $c_-=-(4+8m)$ on the domain wall, thus we obtain the gravitational Chern-Simons theory $(8+16m)\cdot 2 \mathrm{CS}_{\mathrm{grav}}$ on the $T$ domain wall, which carries $c_-=8$ mod $16$.
We note that the bulk action~\eqref{eq:bulkRR} with $c_-=4$ mod $8$ gives the same action as~\eqref{w22} on a closed oriented manifold, since $w_2^2$ is related to the Pontrjagin class as~\cite{Aharony_2013,HsinLorentz}
\begin{align}
    w_2^2=p_1 \ \mod 2,
\end{align}
and then we have
\begin{align}
    \exp\left(\pi i\int w_2^2\right)=\exp\left(\frac{i}{24\pi}\int\mathrm{Tr}(R\wedge R)\right).
\end{align}
We can also understand the chiral domain wall with $c_-=8$ mod $16$ as follows. We denote the $(2+1)$d spacetime for the left domain on the gapped boundary as $X$, see Fig.~\ref{fig:wall}.
The $T$ domain wall in the bulk ends on $\partial X$ at the $(2+1)$d boundary.
Since the $T$ symmetry is preserved on the boundary, $\partial X$ must support a $T$ defect operator of the $(2+1)$d TQFT on the boundary. Because the boundary is a gapped TQFT, the $T$ defect on $\partial X$ must be topological and carry gapped degrees of freedom, which must lead to $c_-=0$ on $\partial X$.

Now, the left and right domain of the TQFT on the boundary contributes $c_-=8$ mod $16$ to $\partial X$, and it must be cancelled by the bulk contribution. Thus, the $T$ domain wall in the $(3+1)$d SPT phase must carry $c_-=8$ mod $16$.
We identify the $c_-=8$ phase on the $T$ domain wall as the $E_8$ state that generates the free part of $\Omega_{\mathrm{SO}}^3(B\mathbb{Z}_2)=\mathbb{Z}_2\times\mathbb{Z}$.
Thus, we conclude that $\beta_2=1$ in the Smith map.
We can further see that $\beta_1=0$, by seeing that the action given by decorating the $(2+1)$d action~\eqref{a3} on the $T$ domain wall evaluates $Z(\mathbb{CP}^2)=1$ since $\mathbb{CP}^2$ is oriented, so cannot generate the action~\eqref{w22}.

Summarizing, the Smith map~\eqref{smith:boson} is given by
\begin{align}
(n_{DW},n_E)\to(n_{DW}, n_E) \quad \mod 2.
\end{align}

\subsection{$(3+1)$d fermionic $T$ SPT phase: $T^2=(-1)^F$}
In the fermionic case, the $T$ symmetry $T^2=(-1)^F$ corresponds to $\mathrm{pin}^+$ structure of the spacetime. The Smith map determines the classification of $(3+1)$d $T$ SPT phase from that of $(2+1)$d $\mathbb{Z}_2$ SPT phase on the $T$ domain wall, expressed as
\begin{align}
    \Omega_{\mathrm{spin}}^3(B\mathbb{Z}_2)\to\Omega_{\mathrm{pin}^+}^4.
    \label{smith:fermion}
\end{align}
This gives a linear map $\mathbb{Z}_8\times\mathbb{Z}\to\mathbb{Z}_{16}$, where the $\mathbb{Z}$ part of $\Omega_{\mathrm{spin}}^3(B\mathbb{Z}_2)$ is generated by the $p+ip$ superconductor with $c_-=1/2$. The $\mathbb{Z}_8$ part corresponds to the $\mathbb{Z}_2$ SPT phase described by the decoration of the Kitaev wire~\cite{Tarantino}.
If we label elements as $(n,k)\in\mathbb{Z}_8\times\mathbb{Z}$ and $\nu\in\mathbb{Z}_{16}$, the map is determined by~\cite{HKT2019CPT} in the form of
\begin{align}
    \nu=2n-k \quad \mod 16.
    \label{smithformulafermion}
\end{align}
In particular, the above formula dictates that the odd $\nu$ must carry odd $k$ on the $T$ domain wall. Namely, $c_-$ of the SPT phase on the $T$ domain wall must be $c_-=1/2$ mod $1$ when $\nu$ is odd, and $c_-=0$ mod $1$ when $\nu$ is even.

\section{Constraint on $(2+1)$d pin$^+$ and bosonic TQFT}
\label{sec:constraint}
We argue that the TQFT on the boundary of $(2+1)$d TQFT has a restricted value of $c_-$, depending on the chiral phase on the $T$ domain wall controlled by the Smith map. For simplicity, we focus on $\mathrm{pin}^+$ anomaly classified by $\mathbb{Z}_{16}$. The generalization to the bosonic case is straightforward.

\begin{figure}[htb]
\centering
\includegraphics{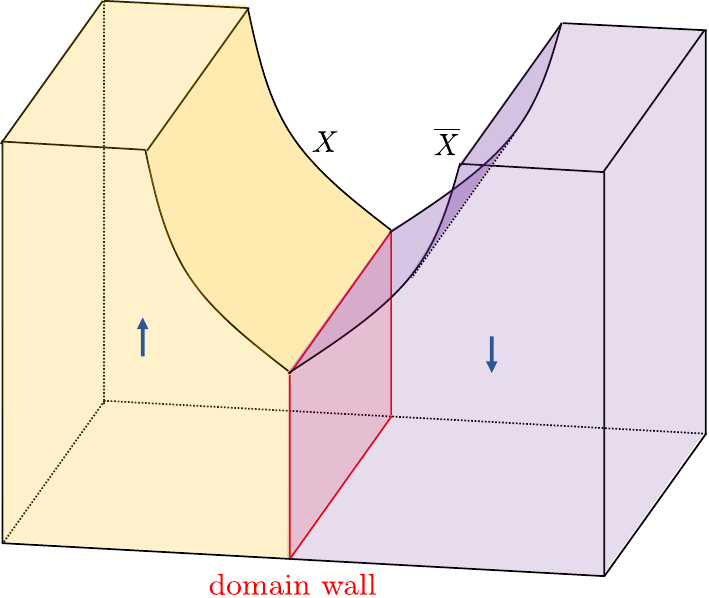}
\caption{The illustration for our setup. We have a $T$ domain wall (red plane) in the $(3+1)$d bulk, which ends at $\partial X$ on the boundary.}
\label{fig:wall}
\end{figure}

Let us consider a $(2+1)$d $\mathrm{pin}^+$ TQFT $\mathcal{T}$ on the boundary of a $(3+1)$d $T$ SPT phase, classified by $\nu\in\mathbb{Z}_{16}$. We again work on the geometry described in Fig.~\ref{fig:wall}, i.e., we break the $T$ symmetry simultaneously in the $(3+1)$d bulk and the $(2+1)$d boundary, such that the $T$ domain wall in the bulk terminates on a domain wall at the boundary. 
If the boundary TQFT on the left domain $X$ has the chiral central charge $c_-=c_{\mathcal{T}}+m/2$ for $m\in\mathbb{Z}$, the right domain $\overline X$ has $c_-=-(c_{\mathcal{T}}+m/2)$ since it is the $T$ partner of the left domain. 
This implies the $(3+1)$d SPT action given by
\begin{align}
    \frac{c_-}{96\pi}\mathrm{Tr}(R\wedge R)
\end{align}
has a kink of $c_-$ from $c_- = c_{\mathcal{T}}+m/2$ to $c_-=-c_{\mathcal{T}}+m/2$ on the $T$ domain wall, thus we obtain the gravitational Chern-Simons theory $(2c_{\mathcal{T}}+m)\cdot 2\mathrm{CS}_{\mathrm{grav}}$ on the $T$ domain wall, which carries $c_-=2c_{\mathcal{T}}$ mod $1$.
Meanwhile, according to the Smith map~\eqref{smithformulafermion}, $c_-$ carried by the $T$ domain wall must be $c_-=\nu/2$ mod $1$. This leads to $c_{\mathcal{T}}=\nu/4$ mod $1/2$.

We can also understand the chiral domain wall with $c_-=1/4$ mod $1/2$ as follows. The $T$ domain wall in the bulk ends on $\partial X$ at the $(2+1)$d boundary. 
Since the $T$ symmetry is preserved on the boundary, $\partial X$ must support a $T$ defect operator of the $(2+1)$d TQFT on the boundary. Because the boundary is a gapped TQFT, the $T$ defect on $\partial X$ must be topological and carry gapped degrees of freedom, which must lead to $c_-=0$ on $\partial X$.
Due to the Smith map~\eqref{smithformulafermion}, the $T$ defect from the bulk contributes as $c_-=1/2$ mod $1$ to $\partial X$, when $\nu$ is odd in $\mathbb{Z}_{16}$. Meanwhile, the TQFT on the boundary contributes $c_-=(c_{\mathcal{T}}+m/2)+(c_{\mathcal{T}}+m/2)=2c_{\mathcal{T}}$ mod $1$.

Thus, in order to have $c_-=0$ on $\partial X$ for odd $\nu\in\mathbb{Z}_{16}$, we must have $2c_{\mathcal{T}}=1/2$ mod $1$, so $c_{\mathcal{T}}=1/4$ mod $1/2$.
For even $\nu\in\mathbb{Z}_{16}$, the $T$ defect instead carries $c_-=0$ mod $1$, so we have $c_{\mathcal{T}}=0$ mod $1/2$.

For the bosonic case, a similar argument shows that $c_-=4$ mod $8$ if the $(2+1)$d TQFT has an anomaly characterized by the SPT action $\int w_2^2$, otherwise $c_-=0$ mod $8$. This is also understood by the fact that $c_-$ mod $8$ is diagnosed by the partition function of the bulk $(3+1)$d SPT phase on $\mathbb{CP}^2$ that corresponds to the Gauss-Milgram formula~\cite{Barkeshli2016}, $Z_{\mathrm{SPT}}(\mathbb{CP}^2)=e^{2\pi i c_-/8}$, which is $-1$ for $\int w_2^2$ and $1$ for $\int w_1^4$.

\section{$(3+1)$d topological superconductor in class AIII}
Here, let us apply our argument to the $(3+1)$d SPT phase with $U(1)\times CT$ symmetry (called class AIII). 
This structure  corresponds to the structure group $\mathrm{pin}^c := (\mathrm{pin}^{\pm}\times U(1))/\mathbb{Z}_2$, where $CT$ corresponds to the orientation reversing element of $\mathrm{pin}^{\pm}$ which commutes with $U(1)$. 

Let us consider the $CT$ defect of the $\mathrm{pin}^c$ $(3+1)$d SPT phase. To see the induced structure on the $CT$ domain wall, it is convenient to regard $\mathrm{pin}^c$ as a $\mathrm{pin}^+$ structure twisted by $U(1)$.
$\mathrm{pin}^+$ induces oriented $\mathrm{spin}$ structure equipped with the $\mathbb{Z}_2$ symmetry on the domain wall, and we also have $U(1)$ symmetry that twists the induced $\mathrm{spin}$ structure. Then, the induced structure on the domain wall becomes $\mathrm{spin}^c$ with $\mathbb{Z}_2$ symmetry.

Therefore, we have the Smith map for cobordism groups
\begin{align}
    \Omega^3_{\mathrm{spin}^c}(B\mathbb{Z}_2)\to\Omega_{\mathrm{pin}^c}^4.
    \label{smithpinc}
\end{align}
The bordism or cobordism groups for these structures are studied in~\cite{GuoPutrovWang,DaiFreed, Gilkeybook}, and given by $\Omega^3_{\mathrm{spin}^c}(B\mathbb{Z}_2)=\mathbb{Z}_4\times \mathbb{Z}\times\mathbb{Z}$, and $\Omega_{\mathrm{pin}^c}^4=\mathbb{Z}_8\times\mathbb{Z}_2$.
We label the elements of $\Omega^3_{\mathrm{spin}^c}(B\mathbb{Z}_2)$ as $(n_{4},n_{CS},n_{E})\in\mathbb{Z}_4\times \mathbb{Z}\times\mathbb{Z}$. The generators are described as follows:
\begin{itemize}
    \item $(0,1,0)$ corresponds to the $\mathrm{spin}^c$ Chern-Simons theory at level $1$, defined via the $(3+1)$d $\mathrm{spin}^c$ $\theta$-term in~\eqref{theta}.
    This theory carries $c_-=1$. 
    \item $(0,0,1)$ corresponds to the $E_8$ state, which carries $c_-=8$.
    \item $(1,0,0)$ generates the $\mathbb{Z}_4$ group, and we believe that it should be formulated in a similar way to the Gu-Wen $\mathbb{Z}_2$ SPT phase based on $\mathrm{spin}$ structure, which is labeled by a pair of cohomological data~\cite{Gu:2012ib, Gaiotto:2015zta}.
    Actually, if we compute the cobordism group by using the toolkit of the Atiyah-Hilzebruch spectral sequence (AHSS)~\cite{DaiFreed}, we see that it can also be described by a pair of cohomological data, which should be regarded as the $\mathrm{spin}^c$ version of the Gu-Wen phase. Namely, the group $\mathbb{Z}_4$ is the nontrivial extension of
    \begin{align}
    H^2(B\mathbb{Z}_2,\Omega_{\mathrm{spin}^c}^1)=H^2(B\mathbb{Z}_2,\mathbb{Z})=\mathbb{Z}_2 
    \end{align}
    by 
    \begin{align}
        H^4(B\mathbb{Z}_2,\Omega_{\mathrm{spin}^c}^{-1})=H^3(B\mathbb{Z}_2,U(1))=\mathbb{Z}_2.
    \end{align}
    So, we expect that the $\mathbb{Z}_2$ subgroup of the $\mathbb{Z}_4$ is given by the bosonic $\mathbb{Z}_2$ SPT phase with the classical action
    \begin{align}
    \exp\left(\pi i \int a^3\right),
    \label{a32}
\end{align}
    
    which characterizes the nontrivial element of $H^3(B\mathbb{Z}_2, U(1))=\mathbb{Z}_2$. 
    Based on the analogy with the Gu-Wen $\mathrm{spin}$ SPT phase, we believe that $H^2(B\mathbb{Z}_2,\Omega_{\mathrm{spin}^c}^1)$ is associated with the physical description that a $(0+1)$d $\mathrm{spin}^c$ SPT phase (in this case a complex fermion with charge $1$) is decorated on the junction of $\mathbb{Z}_2$ defects, and the way for the decoration is controlled by $H^2(B\mathbb{Z}_2,\Omega_{\mathrm{spin}^c}^1)$. 
    Though we have not constructed any action for the $\mathbb{Z}_4$ generator, we believe that there exists a good state sum definition for this theory on lattice like the Gu-Wen $\mathrm{spin}$ SPT phases, which carries $c_-=0$.
\end{itemize}
Meanwhile, if we label the element of $\Omega_{\mathrm{pin}^c}^4$ as $(m_8,m_2)\in\mathbb{Z}_8\times\mathbb{Z}_2$, the actions for the generators are described as follows:
\begin{itemize}
    \item If the spacetime is oriented, the generator $\mathbb{Z}_8$, $(1,0)$ is described by the $\theta$-term for $\mathrm{spin}^c$ gauge field at $\theta=\pi$~\cite{Seiberg2016Gapped, Metlitskiduality}, given by
    \begin{align}
        S[a]=i\theta\left(\frac{1}{2(2\pi)^2}\int f\wedge f-\frac{\sigma}{8}\right),
        \label{theta}
    \end{align}
    where $a$ is a $\mathrm{spin}^c$ gauge field with the Dirac quantization condition
    \begin{align}
        \int_C\frac{f}{2\pi}=\frac{1}{2}\int_C w_2 \quad \mod 1,
    \end{align}
    for any oriented 2d cycle in the spacetime. $\sigma$ denotes the signature of the manifold.
    Also, for later convenience, we note that $m_8=4\in\mathbb{Z}_8$, $(4,0)$ is given by
    \begin{align}
    \exp\left(\pi i\int w_1^4\right).
    \label{w142}
\end{align}
    \item The generator of $\mathbb{Z}_2$, $(0,1)$ is given by
    \begin{align}
    \exp\left(\pi i\int w_2^2\right).
    \label{w222}
\end{align}
\end{itemize}
Then, we can almost completely determine the Smith map~\eqref{smithpinc} by considering a $CT$ domain wall of the $(3+1)$d action. First, since we know in Sec.~\ref{subsec:boson} that the $(3+1)$d action~\eqref{w142} localizes the action~\eqref{a32} on the domain wall, we expect that $(2,0,0)$ is mapped to $(4,0)$ by the Smith map. 
Due to linearity of the Smith map, it shows that $\mathbb{Z}_4$ part of $\Omega^3_{\mathrm{spin}^c}(B\mathbb{Z}_2)$ is mapped to the $\mathbb{Z}_4$ subgroup of the $\mathbb{Z}_8$ part in $\Omega_{\mathrm{pin}^c}^4$. 
According to Sec.~\ref{subsec:boson}, we also know that the $(3+1)$d action~\eqref{w222} for $(0,1)$ localizes the $E_8$ state on the domain wall, so it also determines how $(0,0,1)$ transforms.
Finally, for $(1,0)$ in $\Omega_{\mathrm{pin}^c}^4$, the $CT$ domain wall for the $(3+1)$d $\theta$-term induces a kink of $\theta$ from $\theta=+\pi$ to $\theta=-\pi$, so we obtain a Chern-Simons theory at level $1$ on the domain wall. So, we know how $(0,1,0)$ transforms.
Thus, our Smith map is obtained as
\begin{align}
    (n_4,n_{CS},n_{E})\to(n_{CS}+(2+4p)n_4, n_E),
\end{align}
where $p$ is an undetermined integer.

According to the Smith map, the odd element of $\mathbb{Z}_8$ in $\Omega_{\mathrm{pin}^c}^4$ must carry $c_-=1$ mod $2$ on the $CT$ domain wall. So, by using the same logic as the $\mathrm{pin}^+$ anomaly, we can see that the odd phase in $\mathbb{Z}_8$ must have $c_-=1/2$ mod $1$.

\section{Discussion}
In this paper, we found the anomaly constraint on chiral central charge of $(2+1)$d topological order with $T$ symmetry. The constraint comes from a chiral state localized on the $T$ domain wall in the bulk SPT phase.
It should be interesting to study such a constraint on the $(2+1)$d TQFT enriched by more generic global symmetry, though we have only studied the cases for $T$ and $U(1)\times CT$. For example, by using the AHSS, $(d+1)$d fermionic SPT phases with $G_b$ symmetry is generally labeled by the set of cohomological data~\cite{qingrui,Thorngren2018bosonization}
\begin{align}
    n_p\in H^{p}(BG_b, \Omega_{\mathrm{spin}}^{d+1-p}),
\end{align}
for $0\le p\le d+2$. Here, $G_b$ can contain time reversal, and the full global symmetry is described by $G_f$, defined as the symmetry extension by fermion parity $\mathbb{Z}_2^F\to G_f\to G_b$. 
The data $n_p$ is associated with the description of the SPT phase based on the decorated domain wall; $n_p$ controls the way to decorate an $((d-p)+1)$d SPT phase on the domain wall of $G_b$. 
In particular, the nontrivial $n_1$ implies the decoration of $p+ip$ superconductor on the $G_b$ domain wall. We expect that this description of decorated domain wall leads to a unified formulation of the anomaly constraints on the Hilbert space for a broad class of global symmetries. See also~\cite{delmastro2021global}. 

It is also very interesting to study the constraint on the chiral central charge for crystalline symmetries. In that case,  there is a simple way to reduce the (3+1)d SPT phase to the lower dimensional system with internal symmetries~\cite{Song_2017}. For example, consider a unitary reflection symmetry across the (2+1)d plane which protects the the (3+1)d SPT phase. 
Then, we can operate the unitary circuit respecting the reflection symmetry away from the reflection plane, which can disentangle the SPT phase away from the reflection plane.
After all, we obtain the reduced (2+1)d SPT phase supported on the reflection plane, where the reflection symmetry acts internally. Based on this logic,~\cite{Mao_2020} obtains a similar constraint on chiral central charge for the (3+1)d fermionic SPT phase with the spatial reflection symmetry. 

In the present paper, we worked on the setup with Lorentz invariance and did not discuss the effect of lattice. It is interesting what the Smith map and the global symmetry on the symmetry defects looks like in lattice systems, since we deal with lattice systems in the study of SPT phases which are not Lorentz invariant in general. 
There is a lattice model for the (3+1)d SPT phase with $T^2=(-1)^F$~\cite{Kobayashi_2020}, where in the $T$ broken phase we observe a unitary global $\mathbb{Z}_2$ symmetry on the domain wall, at the level of a lattice model. We believe that this global symmetry reflects the induced $\mathbb{Z}_2$ symmetry via the CPT theorem of the effective field theory. It is interesting to study such a lattice model in more detail, e.g., the gapped boundary of this theory and see how the anomaly constraint works. 

Finally, another interesting direction is to ask if our constraint on $c_-$ is applicable to generic gapless phases, while we have worked on gapped topological ordered states in the present paper. 
We leave these problems to future work.

\section{Acknowledgements}
The author is grateful to Maissam Barkeshli and Kantaro Ohmori for enlightening discussions. 
The author also thanks Yunqin Zheng for helpful comments on the manuscript.
The author is supported by the Japan Society for the Promotion of Science (JSPS) through Grant No.~19J20801.

\bibliographystyle{ytphys}
\baselineskip=.95\baselineskip
\bibliography{ref}

\end{document}